\documentclass[a4paper,twoside,11pt,english]{article}

\usepackage[T1]{fontenc}
\usepackage[latin1]{inputenc}
\usepackage{babel}
\usepackage{amsmath}
\usepackage{latexsym}
\usepackage{graphicx}

\def\beq{\begin{equation}}
\def\eeq{\end{equation}}
\def\bitem{\begin{itemize}}
\def\eitem{\end{itemize}}
\def\bear{\begin{array}}
\def\ear{\end{array}}

\def\teta{\vartheta}

\def\munu{{\mu\nu}}

\def\d{\partial}

\def\dn{\d_{\nu}}
\def\dr{\d_{\rho}}

\def\j{\hat{j}}

\begin{document}


\begin{titlepage} \vspace{0.2in} 

\begin{center} {\LARGE \bf 
Geometry and Matter Reduction in a 5D Kaluza-Klein Framework}\\ 
\vspace*{1cm}
{\bf  V. Lacquaniti $^{1,2,3,\diamond}$, G. Montani $^{1,4,5,\dag}$ }\\
\vspace*{1cm}
$^1$ ICRA---International Center for Relativistic Astrophysics, 
Physics Department (G9),
University  of Rome, "`La Sapienza"', 
Piazzale Aldo Moro 5, 00185 Rome, Italy.\\

$^2$  Physics Department  "`E.Amaldi "`,  University of Rome, "`Roma Tre"', 
Via della Vasca Navale 84, I-00146, Rome , Italy \\ 

$^3$ LAPTH -9, Chemin de Bellevue BP 110 74941 Annecy Le Vieux Cedex, France \\

$^4$ ENEA- C. R. Frascati ( Department F. P. N. ), via E.Fermi 45, I-00044, Frascati, Rome, Italy \\
$^5$ ICRANET - C.C. Pescara, Piazzale della Repubblica 10, I-65100, Pescara, Italy \\

\vspace*{1.8cm}

{\bf   Abstract  \\ } \end{center} \indent
 In this paper we consider the Kaluza-Klein fields equations in presence of a generic 5D matter tensor which is governed by a conservation equation due to 5D Bianchi identities. Following a previous work, we provide a consistent approach to matter where the problem of huge massive modes is removed, without relaxing the compactification hypotheses; therefore we  perform the dimensional reduction either for metric fields and for matter, thus identifying a pure 4D tensor  term, a 4D vector term and a scalar one. Hence we are able to write down 
a consistent set of equations for the complete dynamics of matter and fields ; with respect to the pure Einstein-Maxwell system we now have two additional scalar field: the usual dilaton one plus a scalar source term. Some significant scenarios involving these terms are discussed and perspectives for cosmological applications
are suggested.

\vfill
\hrule

$^{\diamond}$ valentino.lacquaniti@icra.it,
$^{\dag}$ montani@icra.it

\end{titlepage}

\section{The Problem of Matter in Kaluza Klein}

An open issue, in the framework of the compactified 5D Kaluza-Klein ( KK) theory ( \cite{kaluza}-\cite{modernkk} ), is the problem of matter dynamics ( \cite{KK}, \cite{overduinwesson}, \cite{vandongen} ), as well as the presence of the extra scalar field $\phi$ ( \cite{cianfrani}, \cite{witten} ) in the model and its interplay with particles motion.
The common approach to the test particle dynamics consists in a generalization of the geodesic approach we use in General Relativity. Therefore,
given the 5D geodesic equation, $\frac{Dw^A}{Ds_5}=0$, where $ds_5$ is the 5D line element and   $w^A=\frac{dx^a}{ds_5}$ is the 5D velocity, we can consider its dimensional reduction:
the motion equation for the reduced 4D particle reads
\beq
\frac{D}{Ds}u^{\mu}=F^{\munu}u_{\nu}\left(\frac{ekw_5}{\sqrt{1+\frac{w_5^2}{\phi^2}}}\right)+\frac{1}{\phi^3}(u^{\mu}u^{\nu}-g^{\munu})\dn{\phi}\left(\frac{w_5^2}{1+\frac{w_5^2}{\phi^2}}\right)\,,
\eeq
where $w_5$ is the fifth component of the 5D velocity $w^A$ - which is scalar and conserved along the path, due to the Killing vector provided by the cylindricity hypothesis -, $u^{\mu}$ is the 4D velocity of the 4D reduced particle, $\frac{D}{Ds}$ the 4D covariant derivative and $ek$ reads $(ek)^2=\frac{4G}{c^2}$. By mean of the presence of $w_5$, we are able to define the charge-mass ratio of the particle which is  :
 \beq
 \frac{q^2}{4Gm^2}=
 \frac{w^2_5}{1+\frac{w_5^2}{\phi^2}}
 \eeq
 While this procedure seems to be physically well-grounded, it actually  turns into some inconsistencies. The ratio above defined is upper bounded, and we have:
  \beq
  \frac{q^2}{4Gm^2}< \phi^2 
   \eeq
 If we assume $\phi=1$, which is the simplest KK scenario that allows us to recover the electrodynamics, such a bound is always violated ( for the electron case we have $\frac{e^2}{4Gm_e^2}\sim 10^{42}$ ). To overcome such a problem, we could assume a large scalar field $\phi$, but, as noted by some authors ( \cite{oklo}-\cite{oklo4} ), the presence of $\phi$  would affect the definition of the fine structure constant, providing violations of the Free Falling Universality  ( FFU ) of particles in electromagnetic bounded systems. In general, however, a large value for the scalar field, which is the scale factor of the extra dimension, would require a fine tuning on the coordinate length of the fifth dimension, in order to ensure its unobservability, or otherwise it would lead us to the framework of large extra dimension model. The problem of the correct evaluation of the charge-mass ratio is indeed a longstanding puzzle in the framework of the compactified model. It is strictly linked, furthermore to the problem of the huge massive KK  modes. It is possible to show, via an hamiltonian analysis, that the charge is directly proportional to $P_5$, i.e. the fifth, conserved,  component of the 5D momentum. Taking into account the compactification of the extra dimension, we get a discrete spectrum of modes for the momentum, thus explaining the quantization of charge; fixing the ground value of the spectrum via the elementary charge $e$ we get an extra dimension length below our observational bound, but, given the link between charge and mass, we obtain a huge massive modes spectrum beyond Planck scale.
Therefore this approach is not consistent with Lorenzian dynamics.
This outcome is a direct consequence of the fact that we adopted a geodesic approach to describe the motion of the unreduced 5D test particle in our compactified model. Other approaches to the test particles dynamics ( and to extra dimensional physics in general ) adopt  different schemes concerning the properties of the extra dimension, adopting a projective approach ( \cite{projective}-\cite{pro4} ) or relaxing the cylindrical and the compactification hypotheses, like in brane-models ( \cite{brane}, \cite{brane2} ) or induced matter theory ( \cite{overduinwesson} ),
and deal with non compact space and embedding procedure ( \cite{rub}-\cite{shap2}  ) -which at the end leads to the brane scheme-, or still adopt the compactification hypothesis but face the problem of the motion using more than 5D and invoking a symmetry breaking mechanism for the generation of mass ( \cite{collins} ).
 In this report we would add a proposal to this debate, by considering an approach where, rather than giving up the compactification hypothesis or the 5D background, we criticise the assumption that the 5D motion is a geodesic one and we start our analysis by analysing the full field dynamics involving a generic 5D tensor source.


\section{Revised Approach to Matter}
In a recent paper ( \cite{KK}), it has pointed out that the geodesic approach is not the proper method to face the motion of a test particle in the compactified 5D KK background, because it does not fulfils two fundamental requests that actually hold when we consider the analogue procedure in GR. They are the Equivalence Principle, which is indeed violated in the 5D framework, and the assumption that we can define the 5D particle as a 5D point like object, which is no more guaranteed by the  compactification of the fifth dimension at a low scale such that we cannot probe it at the present allowable energy scale. It has actually been suggested that the proper procedure is to deal with a 5D conservation law concerning a generic  5D matter, i.e. $^5\nabla_AT^{AB}=0$, and achieve the definition of the particle trough a multipole expansion, as it happens in GR with the Papapetrou procedure ( \cite{papapetrou}).\\
Here we follow such an approach, embedding it in the general point of view of the full KK fields dynamics; the aim is to enforce it, providing a physical grounded meaning to the tensor $T^{AB}$, and to achieve a consistent scheme for a complete reduction of matter and geometry.\\
\subsection{Dynamics in vacuum}
It is worth starting our analysis by recalling the formulation of the KK dynamics in vacuum.
Let us consider at first the KK model in vacuum. The dynamics of fields is given by 5D Einstein equations $ ^5G_{AB}=0$ or - what is not trivial to be remarked, being in principle not guaranteed- by the variational procedure applied on the 5D Einstein-Hilbert Action which reads, in $c=1$ units, $S_5=-\frac{1}{16\pi G_{5}}\int\!\!d^5\!x \sqrt{J}\,\, {}^5\!R
$, where $J^{AB}$ is the 5D metrics, ${}^5\!R$ its associated curvature scalar and $G_{5}$ is an unknown 5D Newton constant. Performing the dimensional splitting, and after some  rearrangements, both procedure lead to  the  set ( \cite{jordan}, \cite{thirry} ):
\begin{eqnarray}
&{}&G^{\munu}= \frac{1}{\phi}\nabla^{\mu}\d^{\nu}\phi-\frac{1}{\phi}g^{\munu}\Box\phi+8\pi G\phi^2T^{\munu}_{electrom.} \nonumber \\
&{}&\nabla_{\mu}\left(\phi^3 F^{\munu}\right)=0 \nonumber \\
&{}& \Box\phi=-\frac14\phi^3(ek)^2F^{\munu}F_{\munu} \, .
\label{kkeqvacuum}
\end{eqnarray}
Here the usual Newton constant $G$ naturally arises defined as 
$$
\frac{1}{G}=\frac{1}{G_5}\int\!dx^5\, ,
$$
where $l_5=\int\!dx^5$ is the coordinate length of the extra dimension.\\
In principle,  we would like to consider the scenario $\phi=1$, in order to restore exactly the Einstein-Maxwell dynamics. However,  such a scenario provides an ambiguous feature: first two equations reduce to Einstein-Maxwell dynamics, while the last turns into the inconsistent result $F_{\munu}F^{\munu}=0$. Therefore, the hypothesis $\phi=1$ can be considered only if we request it on the Action, before doing the variational procedure that leads to equations ( \ref{kkeqvacuum} ).
\subsection{Reduction of matter}
To introduce matter in the dynamics, we now assume the existence of a 5D matter tensor $\mathcal{T}^{AB}$ and therefore we write down the full system for the 5D Einstein equation in presence of 5D matter:
\beq
^5G^{AB}=8\pi G_5 \mathcal{T}^{AB}
\label{kkk}
\eeq
where $G_5$ is the 5D Newton constant.
Such an equation could be derived from the 5D Lagrangian density
$$
L_5= -\frac{1}{16\pi G_5}\sqrt{J}{\,}^5\!R+\sqrt{J}{\,}^5\!\mathcal{L}
$$
where the first term is the 5D Einstein-Hilbert Lagrangian and the second is the  Lagrangian for the 5D matter.
Our 5D matter tensor is defined in term of the Lagrangian density as follows:
$$
\mathcal{T}^{AB}=2\left( \frac{\delta{}^5\mathcal{L}}{\delta J^{AB}}-\frac12 J_{AB}{}^5\mathcal{L}    \right )
$$
It is worth remarking at this step that we consider the 5D Lagrangian density and the 5D matter tensor as unknown objects.  The dimensional reduction eventually will allow us to link the degrees of freedom of the unknown 5D tensor to some recognisable 4D objects,  therefore allowing us to provide their physical interpretation.
At first, we note that the components of $\mathcal{T}^{AB}$ should represent a energy density with respect to the spatial volume , which is a 4D volume including the extra dimension; hence, in order to move toward a 4D tensor, which is related to energy density with respect to a 3D spatial volume,  we need at first a rescaling via a proper length scale. Therefore,
given the coordinate length of the extra dimension, $l_5=\int dx^5$, we define $T^{AB}=\mathcal{T}^{AB}l_5$, and at the same time we achieve the rescaling $G=G_5l_5^{-1}$ in the equation  ( \ref{kkk} ) which is the same reparametrization we have in vacuum.
As a second step, we observe that is possible to demonstrate that the 5D reduced Bianchi identities hold, therefore the divergence of the tensor vanishes, giving us a conservation equation, as a consequence of the 5D invariance with respect to traslations.
Finally, being metrics field governed by the cylindricity hypotheses, we also assume that such an hypothesis holds for the matter tensor.

At last,  we deal with the following model:
\begin{eqnarray}
\label{bianchi5d}
&{}& ^5\nabla_A T^{AB}=0 \\
\label{einstein5d}
&{}& ^5G^{AB}=8\pi G {T}^{AB} \\
\label{cyl}
&{}&
\d_5T^{AB}=0
\end{eqnarray}
First two equations represent the proper 5D generalization of the 5D Einstein equation, while the last takes into account the cylindricity hypothesis concerning matter fields.
The dimensional reduction of eq. (\ref{bianchi5d}), after some algebraic rearrangements, yields the following set:

\begin{eqnarray}
\label{bianchi5}
&5)& \rightarrow \nabla_{\mu}\left(\phi T_5^{\mu}\right)=0 \quad 
\\
\label{bianchimu}
&\mu)&\rightarrow \nabla_{\rho}(\phi T^{\mu\rho})=-g^{\mu\rho}\left(\frac{\dr\phi}{\phi^2}\right)T_{55}+ ekF^{\mu}_{\,\,\rho}\phi T_5^{\rho}
\end{eqnarray}
It is worth noting that the components $T^{\munu}$, $T_5^{\mu}$, $T_{55}$ are a 4D tensor, a 4D vector and a scalar respectively.
It looks like convenient to define:
\beq
T^{\munu}_{matter}=\phi T^{\munu} \quad\quad j^{\mu}=ek\phi T_5^{\mu}
\label{redef}
\eeq
This way, we have:
\begin{eqnarray}
\label{corrcons}
&{}& \nabla_{\mu}j^{\mu}=0\\
\label{cons}
&{}&\nabla_{\rho}( T^{\mu\rho}_{matter})=-g^{\mu\rho}\left(\frac{\dr\phi}{\phi^2}\right)T_{55}+ F^{\mu}_{\,\,\rho}j^{\rho}
\end{eqnarray}
Therefore, the first equation gives us a conserved current $j_{\mu}$, coupled to the tensor $F_{\munu}$, while the second describes the dynamics of the tensor $T^{\munu}_{matter}$, which is coupled to the field $\phi$ and $A_{\mu}$ trough the matter terms $T_{55}$ and $\j_{\mu}$.
We thus interpret $j^{\mu}$ as the proper current related to the U(1) gauge symmetry and $T^{\munu}_{matter}$ as the proper tensor representing the energy-momentum density of the ordinary 4D matter. The last statement is enforced observing that setting $\phi=1$ we recover the conservation law for an electrodynamics system, while, setting as zero the other source term $T_{55}$ and $j^{\mu}$ we recover the law $\nabla T^{\munu}_{matter}=0$, which is what we expect from an  isolate 4D matter system.
As well as we have done for the tensor and the vector component, we also re-scale the scalar component defining now $\teta=\phi T_{55}$.
A generic heuristic argument can be provided as a support to this kind of scaling in term of $\phi$.
Let us  compare the definition of the starting tensor $\mathcal{T}^{AB}$ with respect to the definition of  the matter tensor $T^{\munu}_{matter}$ in GR;  as previously argued,  the hypothetic 5D observer, will perceive a 4D spatial volume, including the contribute of the extra dimensions, therefore we  will have these two definitions, being $x$ a generic label for the components of the tensor and $E^{x}$ an object with the dimension of an energy:
\begin{eqnarray}
5D  &\rightarrow& E^x=\int dx^5d^3x \sqrt{J}\mathcal{T}^{x} \\
4D  &\rightarrow& E^x=\int d^3x \sqrt{g}T^{x}_{matter}
\end{eqnarray}
Now, using the cylindricity hypothesis , the compactification, and the reduction formula $\sqrt{J}=\phi\sqrt{g}$ we have:
\beq
5D  \rightarrow E^x=\int d^3x \sqrt{g} \left(\phi l_5 \mathcal{T}^{x}\right)  
\eeq
Therefore, a comparison between the above equations suggests  to define, for instance,
$$
T^{\munu}_{matter}= l_5 \phi  \mathcal{T}^{\munu}=\phi T^{\munu}
$$
which is indeed the definition we naturally adopted following the dimensional procedure.

Taking now into account these definitions, the reduction of eq. (\ref{einstein5d}) leads to the set
\begin{eqnarray}
\label{tens}
&{}&  G^{\munu}=\frac{1}{\phi}\nabla^{\mu}\partial^{\nu}\phi-\frac{1}{\phi}g^{\munu}\Box\phi+8\pi G\phi^2T^{\munu}_{electrom.}+8\pi G \frac{T^{\munu}_{matter}}{\phi} \\
\label{vec}
&{}&  \nabla_{\nu}\left(\phi^3F^{\nu\mu}\right)=4\pi j^{\mu} \\
&{}&   \frac12 R+\frac38 \phi^2(ek)^2F^{\munu}F_{\munu}=8\pi G \frac{\teta}{\phi^3} \nonumber
\end{eqnarray}
Arranging the last equation with the trace of the first we have:
\beq
\Box\phi=-\frac14 \phi^3(ek)^2 F^{\munu}F_{\munu}+\frac83 \pi G \left( T_{matter}+2\frac{\teta}{\phi^2}\right)
\label{scal}
\eeq

 This set of equations is supported by eq. ( \ref{bianchimu} ) which now we rewrite in terms of new variables:
 \begin{eqnarray}
\label{corrcons2}
&{}& \nabla_{\mu}j^{\mu}=0\\
\label{cons2}
&{}&\nabla_{\rho}( T^{\mu\rho}_{matter})=-g^{\mu\rho}\left(\frac{\dr\phi}{\phi^3}\right)\teta+ F^{\mu}_{\,\,\rho}j^{\rho}
\end{eqnarray}
At first, we note now that the problem concerning the scenario $\phi=1$ is  removed. In such a scenario eq. ( \ref{cons2} ) leaves $\teta$ undetermined, eq. ( \ref{tens}, \ref{vec} ) describe a Einstein-Maxwell system, and finally eq.  ( \ref{scal} ) turns into a condition which fixes $\teta$ for a given background. Hence, it is possible to consider a scenario for metric fields where, assuming $\phi=1$, we get the expected coupling between matter and fields rebuilding exactly the Einstein-Maxwell dynamics.\\
In general, we conclude that equations ( \ref{tens}, \ref{vec}, \ref{scal} ,\ref{corrcons2}, \ref{cons2})  describe the full KK dynamics in presence of matter source terms.\\
At the end of the dimensional reduction procedure, the model looks like a modified gravity theory where the modification has to be addressed to the presence of the two extra degrees of freedom we have in the fields sector and in the matter sector respectively.
It is important to emphasise that the equation ( \ref{cons2} ), which we have derived from the 5D conservation law induced by the 5D Bianchi identities, could be as well derived letting the operator $\nabla_{\mu}$ act on the equation ( \ref{tens} ) and using the 4D Bianchi identity, although the proof is very long and consists in a tricky succession of algebraic steps. It is important however  to stress that this result enforces the consistency of the whole  procedure. In the same way, finally, eq. ( \ref{corrcons2} ) is a direct consequence of eq. ( \ref{vec} ).

\paragraph {Test particles dynamics}
 The analysis concerning the dynamics of test particles has been done in ( \cite{KK} ), starting from eq. (\ref{cons}) and adopting a Papapetrou multipole expansion , under the hypothesis that the test particle is described by a localized source. We just recall now the outcome for ease of completeness.
 The particle turns out to be delocalized into the fifth dimension, and the 
  equation describing the 4D motion  reads:
 \beq
m\frac{Du^{\mu}}{Ds}=A(u^{\rho}u^{\mu}-g^{\mu\rho})\frac{\d_{\rho}\phi}{\phi^3}+qF^{\mu\rho}u_{\rho}
\label{eqmotion}
\eeq
The scalar coupling factors are defined as follows:
\beq
\!\! m\!=\!\frac{1}{u^0}\!\!\int\!\!\!d^3x\,\sqrt{g}\, T^{00}_{matter} \,\,\,
q\!=\!\!\int\!\!\!d^3x\,\sqrt{g}\, {j^0} \quad
 A\!=\!u^0\!\!\int\!\!\!d^3x\,\sqrt{g}\, \teta
\eeq
 It turns out that now $m$ and $q$ are  not correlated via $w_5$; they are defined in terms of independent degrees of freedom ( $T^{00}$ and $T_5^0$ ) and  therefore we have no bound on $q/m$. The charge is conserved, as required by the presence of the $U(1)$ symmetry; A is not conserved but there is no symmetry that requires it. Actually, the relevant feature is that the mass term now is not conserved ; its behaviour is given by the condition\\
\beq
\frac{dm}{ds}=-\frac{A}{\phi^3}\frac{d\phi}{ds}
\label{massvar}
\eeq
Such a condition arises during the Papapetrou procedure in order to satisfy the requirement $u_{\mu}\frac{D}{Ds}u^{\mu}=0$.
Finally, it is possible to show, via the Hamiltonian formulation of the dynamics and the canonical quantization procedure, that, as well as the $q/m$ puzzle is removed, the tower of huge massive modes now disappear by the presence of an appropriate counter-term in the dispersion relation.

\section {Simple scenarios}
Putting together the fields dynamics with the test particles dynamics we now consider some interesting scenario  which in our opinion deserve some effort to be pursued.
For sake of simplicity we assume from now on $j_{\mu}=0$, $F_{\munu}=0$;  the equations we are interested in 
are (\ref{scal}, \ref{eqmotion}, \ref{massvar}) :

$$
\Box\phi=\frac83 \pi G \left( T_{matter}+2\frac{\teta}{\phi^2}\right)
$$
$$
\frac{dm}{ds}=-\frac{A}{\phi^3}\frac{d\phi}{ds} \quad\quad m\frac{Du^{\mu}}{Ds}=A(u^{\rho}u^{\mu}-g^{\mu\rho})\frac{\d_{\rho}\phi}{\phi^3}
$$
\bitem
\item
$2\teta=- \phi^2T_{matter}$ \\
This is the simplest case: we have as a suitable solution of the scalar equation $\phi=1$; therefore we recover $m=cost$ and $\frac{Du^{\mu}}{DS}=0$, regardless the presence of $A$, which now does not affects the motion equation, thus  we rebuild the Free Falling Universality ( FFU ) of the particle. Therefore we have  the   GR  theory \\
\item
$\teta=0 $ \\
Within this scenario we have $m=cost$, being $A=0$, therefore we have again FFU and the equation $\frac{Du^{\mu}}{DS}=0$, but in principle $\phi$ is now variable and we can look for effects of physics beyond GR. \\
\item
$A=\alpha m \phi^2$ \\
In such a case we still have FFU, because $m$ is ruled out from the motion equation, but in general we have  $\phi$ variable as well as $m$.
Noticeably, equation ( \ref{massvar} ) can be now easily integrated and we have a scaling law for the mass:
\beq
m=m_0\left(\frac{\phi}{\phi_0}\right)^{-\alpha}
\label{scalingmass}
\eeq
It is worth noting that similar equations are interesting for various model of dark matter ( \cite{anderson} ) coming out from modified gravity scenarios.
 \eitem
 Other interesting hints can be achieved by considering some simplified backgrounds like for instance the case of homogeneous or spherically symmetric backgrounds.
At first, it is possible to prove that, as far as the homogeneous scenario is concerned, the comoving velocity $U^{\mu}= (1,0,0,0)$ is still a solution of the motion equation, although the motion equation is modified by presence of scalar fields. This allows us to still employ the definition $T^{\munu}_{matter}=(\rho+p)U^{\mu}U^{\nu}-g^{\munu}p$ for a perfect fluid described in terms of its pressure and density, and suggests to consider a simple parametrization of the extra source $\teta$ like $\teta=\phi^k\left(\alpha \rho +\beta p\right)$. Such a parametrization mimics the equation of state $p=\gamma\rho$ and at the same time takes into account the presence of $\phi$. It is worth noting, moreover, that the case $k=2$ yields, for a dust matter,  to the scenario $A=\alpha m \phi^2$ we discussed above.
On the other side, looking at the exterior solution in a spherically symmetric background ( Generalized Schwarzschild Solution, \cite{sorkin}-\cite{wessonsolDM} ) it is worth remarking that the behaviour of the mass distribution is given in the exterior region by an equation of the form
 $$M_g(r)\, \infty  \,\phi^{-1},$$
 which thus coincides to our equation ( \ref{scalingmass} ) for $\alpha=1$.


\section*{Concluding Remarks}

The usual approach to matter in KK starts from the particles dynamics and results in a failure when compared to the phenomenology, as far as the compactification hypotheses is taken into account. In this report we tried a different approach, starting from the general dynamics of fields and it turns out that this scheme offers a scenario to deal consistently with matter in the framework of the cylindrical and compactified KK model.
 A key point is the extension of the cylindricity hypothesis to the matter tensor, which becomes localized in 4D, i.e. the test particle is not localized in the extra dimension. 
 Therefore the picture of the compactified model we discussed in this work represents a new point of view, with respect to the problem of matter in Kaluza-Klein theories, and in our opinion it deserves further investigations.
If we restrict to the minimal scenario $\phi=1$, which is now allowed as a suitable solutions of the reduced 5D Einstein equations, then we reproduce the Einstein-Maxwell dynamics, either for fields, either for matter, providing consistent definitions for the involved coupling factors. With respect to such a scenario, the Kaluza-Klein model should be considered as a valuable toy-model in order to develop a unifying geometrical picture of interactions. In this sense, it should be stressed that the revised approach we discussed allows us to keep in the extradimensional scenario the presence of an extra dimension compactified at a very low scale; this is relevant, because it provides us an elegant explanation for the discretization of the charge and for the generation of the local U(1) gauge symmetry. With respect to this scenario, the perspective of an extension of this approach to multidimensional models with more than five dimension should be considered in order to face the coupling of the other interactions.
As far as just the 5D model and its phenomenological implications are considered, however, the most promising perspectives involve the presence of  scalar fields in the dynamics.
It could be suggested that a decreasing behaviour of the field $\phi$ could be responsible of the shrink of the fifth dimension, giving a close link between a suitable compactification scenario and the delocalization of particles within the extra dimension; with this respect, an analysis of the suitable behaviours of the scalar field in presence of matter should be recommended.
A natural arena where is possible to test the outcomes of this model is the framework of modified gravity theories. Indeed, among various approaches to the dark matter/energy puzzle, a common one is to deal with modification addressed to the presence of scalar field ( \cite{wetterich}-\cite{gasp} ). As far as the model we just presented here is concerned, a natural path of investigation should be to address the field $\teta$ to a dark matter source, and the field $\phi$ to dark energy. 
With this respect, it appears interesting to develop the model in some simplified scenario, like the framework of an homogeneous background or spherical symmetry. A comparison with theory of extended gravity indeed, would give insight about the possibility of this model to deal with dark matter scenario and provide constraint on the scalar degree of freedom ( which in principle could be ruled out ). It can be stressed, for instance, that the model we discussed is able to provide a physical ground for the existence of modified spherical solutions, which otherwise, in vacuum, should be considered solitons or naked singularities ( \cite{sorkin}-\cite{wessonsolDM} ), and, in more detail, to provide a scenario for the search of modified black holes ( \cite{chardonnet} ).
It is worth noting, finally, that at the end of the work we outlined some scenarios where, although some modified effects are still taken into account, the Free Falling Universality of particles is recovered.
Among these minimal scenarios,  the most promising one is that where we chose to parametrize the extra scalar source in terms of a power of the field $\phi$ times a linear combination of density and pressure;  depending on the parameter of this picture, such a representation fulfils the Free Falling Universality of the particles motion,
and  it therefore appears as a good candidate for a physical check of the model, either with respect to a Friedmann/Schwarschild-like framework either for a study involving perturbations with respect to an assigned homogeneous background.

\section*{Acknowledgements}

The work of V. Lacquaniti has been  supported by a fellowship "Bando Vinci" granted from the "France-Italy University" and has been mostly developed during a stage at LAPTH. V. Lacquaniti thanks Prof. Pascal Chardonnet ( Lapth-University of Savoie ) and Prof. Orlando Ragnisco (  RomaTre ) for many interesting discussions.

\end{document}